# ORGANIZATIONAL PROPERTIES OF SECOND-ORDER VISUAL FILTERS SENSITIVE TO THE ORIENTATION MODULATIONS

Babenko V.V., Miftakhova M.M., Yavna D.V.

*mbmiftahova@sfedu.ru*

*Here we show psychophysical study of second-order visual mechanisms sensitive to the orientation modulations. Selectivity to orientation, phase and spatial frequency of modulation is measured. Bandwidths for phase ($±0,5π$) and orientation ($±33,75$ deg) are defined, but there is no evidence for spatial frequency selectivity.*

Studying preattentive mechanisms of spatial grouping of local information functioning at early stages of visual processing and known as second-order mechanisms is necessary for understanding how local features filtered on the level of primary visual cortex are united into cognitive blocks.

Generally to describe principles of second-order stimuli processing filter-rectify-filter model is used ([1] for review). Firstly this model was non-specific to the modulated parameter (contrast, orientation or spatial frequency (SF)).

Thus, if contrast of texture elements lying in the periphery of the second-order receptive field decreases, then second-order filter's response increases. But we come to the similar results if orientation or SF of elements varies. In other words second-order filters pass any modulation. Kingdom et al. *put this non-specificity into question* for the first time in 2003 [2]. His research carried out in the paradigm of subthreshold summation and the number of latter studies using masking [3] and adaptation paradigms consider the existence of independent channels, selectively sensitive to the *type* of second-order information.

Previously we have shown properties of the second-order visual mechanisms sensitive to the contrast [4] and SF [5] modulations. This research is devoted to the measuring tunings of second-order filters sensitive to the orientation modulations.

**Procedure**
3 observers with normal or corrected-to-normal vision participated in the research.
Experiment was carried out using masking, two-alternative forced-choice procedure, and staircase method.
There were 3 experimental series: measuring selectivity to orientation, SF and phase of modulation. Textures composed of gabor micropatterns were used as stimuli. Amplitude of modulation threshold in the presence of masking stimuli was measured in all experiments.

In each experimental series 5 masks were presented. In the first experiment axis tilt was varied in a range of 0-90 dergees with 22.5 increment from the test stimulus. In the second experiment masks with phase shift from 0 to $1π$ were used, $0.25π$ increment. And in the third experimental series SF of the modulation was changed from -2 to 2 octaves with 1 octave increment. In the control series non-modulated texture was used as a mask.

**Selectivity to the orientation of modulation**
Studying selectivity of second-order mechanism to the orientation of modulation revealed significant decrease of threshold amplitude with increasing of the axis tilt of mask's modulation (Fig. 1)

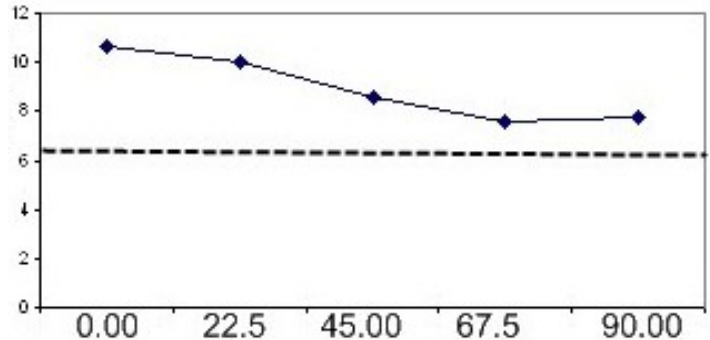

Fig. 1. Threshold amplitude dependence on the orientation of modulation. Summary data for 3 observers. Abscissa shows modulation axis tilt in mask in deg, ordinate shows measured threshold in arbitrary units. Dash-line corresponds to the threshold level when masking by non-modulated mask.

Curves for all observers were similar. Maximum of the masking effect was observed when axis tilts of modulation in mask and test where the same. With the increasing of the difference between test and mask significant threshold decrease was measured. Threshold for masking by 67.5 and 90 degrees axis tilts is about the same and equals 7.55 and 7.74 arbitrary units. Bandwidth at the half of the amplitude is estimated at ±33.75 degrees.

It should be noted that bandwidth for orientation of second-order mechanisms sensitive to the orientation modulations is narrower than bandwidth for second-order mechanisms sensitive to the contrast modulations [4].

**Selectivity to the phase of modulation**
When masking by stimuli with shifted phase of modulation the following change in threshold was found (Fig.2):

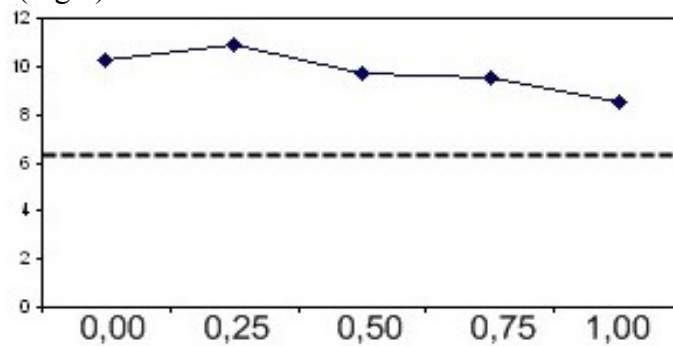

Fig.2. Threshold amplitude dependence on the phase shift of modulation. Summary data for 3 observers. Abscissa shows phase shift of masks's modulation function (parts of $\pi$), ordinate shows threshold in arbitrary units. Dash-line corresponds to the threshold level when masking by non-modulated mask.

It was observed that modulation threshold amplitude decreases with the increase of phase shift of mask's modulation. Therefore we consider second-order mechanisms sensitive to the orientation modulations are sensitive to the phase shift of modulation.

In this experimental series 2 observers have shown bandpass tuning, but one didn't show this tuning. One explanation is subjective difficulty of performing task for this observer, another explanation is that mechanisms that lie higher than second-order mechanism influence on the task performing.

Tuning of filters to the phase of the envelope is estimated as $0.5\pi$ at the half of amplitude, and it is consistent with analogous bandwidth of mechanisms sensitive to the contrast modulations.

**Selectivity to the spatial frequency of modulation**

Studying selectivity to SF of modulation have shown the increase of threshold amlitude with increasing of mask's SF from -2 to +2 octaves (Fig.3).

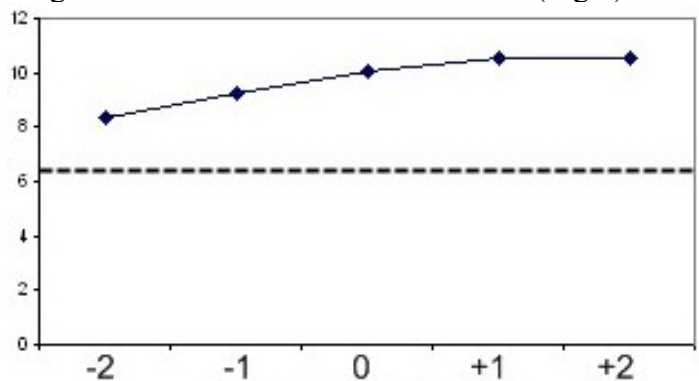

Fig.3. Threshold amplitude dependence on the SF of modulation. Summary data for 3 observers. Abscissa shows SF of mask in octaves, ordinate shows threshold in arbitrary units. Dash-line corresponds to the threshold level when masking by non-modulated mask.

When SF of test and mask are the same threshold amplitude equals 10.5. With decreasing of mask's SF perception of test stimulus improves; with increasing of SF threshold slightly rises.

Results for two observers have shown that detection threshold is significantly lower for low mask's frequencies versus thresholds for high mask's frequencies.

However our data doesn't lead to the conclusion about second-order SF sensitivity of mechanisms sensitive to the orientation modulations.

**Discussion**

Our data allows us to estimate tunings of filters sensitive to the orientation modulations to orientation (±33,75 deg) and phase (±0,5 π) of modulation.

In our previous studies we investigated second-order mechanisms sensitive to the contrast [4] and spatial frequency [5] modulations. According to the data on selectivity we suggested possible forms of receptive fields of underlying second-order vision mechanisms. Receptive fields of mechanisms sensitive to the contrast modulations are likely to have elongated shape and inhibitory subfields. Second-order mechanisms sensitive to the spatial frequency modulations are likely to have concentric receptive fields.

Our current data doesn't allow us to assess SF tunings of second-order mechanisms sensitive to the orientation modulations, because existence and properties of spatial frequency selectivity are not clear. So lack of spatial frequency selectivity can be explained by organizational properties of second-order filters sensitive to orientation modulations. Generally our data is consistent with conclusions of Hallum et al. [6] about the existence of differences in mechanisms processing second-order orientations in the processing of high and low spatial frequencies.